# Thermal camera based on frequency upconversion and its noise-equivalent temperature difference characterization

Zheng Ge,[1, 2] Zhi-Yuan Zhou,[1, 2, *] Jing-Xin Ceng,[3] Li Chen,[1, 2] Yin-hai Li,[1, 2] Yan Li,[1, 2] Su-Jian Niu,[1, 2] and Bao-Sen Shi [1, 2, *]

[1] *CAS Key Laboratory of Quantum Information, University of Science and Technology of China, Hefei, Anhui 230026, China*

[2] *CAS Center for Excellence in Quantum Information and Quantum Physics, University of Science and Technology of China, Hefei 230026, China*

[3] *Science and Technology on Electro-Optical Information Security Control Laboratory, Tianjin 300308, China*

[*] *zyzhouphy@ustc.edu.cn*

[*] *drshi@ustc.edu.cn*

We present a scheme for estimating the noise-equivalent temperature difference (NETD) of frequency upconversion detectors (UCDs) that detect mid-infrared (MIR) light. In particular, this letter investigates the frequency upconversion of a periodically poled crystal based on lithium niobate, where a mid-infrared conversion bandwidth of 220 nm can be achieved in a single poled period by a special design. Experimentally, the NETD of the device was estimated to be 56 mK for a mid-infrared radiating target at a temperature of 95°C. Meanwhile, a direct measurement of the NETD was performed utilizing conventional methods, which resulted in 48 mK. We also compared the NETD of our UCD with commercially available direct mid-infrared detectors. Here, we showed that the limiting factor for further NETD reduction of our device is not primarily from the upconversion process and camera noise, but from the limitations of the heat source performance. Our detectors have good temperature measurement performance and can be used for a variety of applications involving temperature object identification and material structure detection.

Spectroscopy in the mid-infrared band has long been of great value in fields such as environmental monitoring [1–3], biomedicine [4–6], communications [7,8], and remote sensing [9]. This band is closely related to the thermal radiation of the object and contains the absorption/emission spectral positions of numerous molecules and structures [10]. Despite a long history of research on the nature and application of mid-infrared light, the development of the corresponding detectors is still unsatisfactory. Compared with their visible or near-infrared (NIR) counterparts, mid-infrared detectors also suffer from low detection sensitivity, high noise, and narrow bandwidth. In addition, due to the inherent thermal noise of low bandgap materials, such detectors often rely on deep cooling, which imposes an additional burden on the application. The use of high-performance detectors based on wide-bandgap materials (e.g., silicon) to detect mid-infrared light after frequency conversion to visible/near-infrared light has proven to be an effective alternative [11–14]. This technology has been rapidly developed in recent years, with better conversion efficiencies achieved using waveguides [15,16], pulsed light [17,18], and cavity enhancement schemes [19,20]. However, the upconversion process also introduces other noises into the detection results, so it is critical to analyze and evaluate the noise performance of the device. In previous related work, the focus has often been on the evaluation of noise equivalent power [21–24]. This parameter is applicable to areas such as spectral detection, but is not intuitive enough for thermal imaging applications. The noise equivalent temperature difference (NETD) is an important measure of the noise performance of conventional thermal imagers and is defined as the equivalent temperature difference between the target and the background when the signal-to-noise ratio of the image signal is 1. Though being an important parameter for thermal cameras, this parameter has not been theoretically and experimentally studied in previous works for UCD. This work gives the first NETD evaluation and calculation based on frequency upconversion thermal imagers and demonstrates the good noise performance of our experimental setup.

Upconversion imaging is based on sum frequency generation (SFG) in nonlinear crystals, where mid-infrared signal light with frequency $\omega_s$ is upconverted to $\omega_{up}$ by pump light with frequency $\omega_p$. This process satisfies the law of energy conservation, i.e., $h\omega_{up} = h\omega_s + h\omega_p$, where h is Planck's constant. In order to achieve the highest frequency conversion efficiency, Quasi-phase-matching (QPM) techniques are often used to compensate for the phase mismatch and the polarization period is designed such that $\Delta k = 0$, satisfying the momentum conservation condition. Currently, periodically poled lithium niobate (PPLN) crystals have been more widely used to achieve infrared upconversion imaging [25], but single-period crystals usually suffer from a narrow conversion bandwidth. Tuning the temperature [26], scanning the pump wavelength [27], and rotating the crystal angle can extend the conversion wavelength range, but significantly increases system and measurement complexity. Another solution is the use of chirped crystals, which have been used to achieve adiabatic nonlinear conversions with large phase-matching bandwidths [15,28,29]. However, this solution also has stringent requirements for crystal design processing and pumped optical power. In this letter, a broad spectral conversion of the mid-infrared beam can be achieved using only single-period PPLN crystals in a specific wavelength band, using a similar approach to previous work[30]. The variation of the phase mismatch within the crystal with the wavelength of the signal light is calculated in Fig. 1(b) for fixed experimental parameters. The line shape in the figure has an inflection point at $\lambda = 4.14$ μm, where the rate of change of $\Delta k$ is small within a certain bandwidth in the vicinity, and thus frequency conversion can be achieved for bandwidths above 200 nm, as shown in Fig. 1(c).

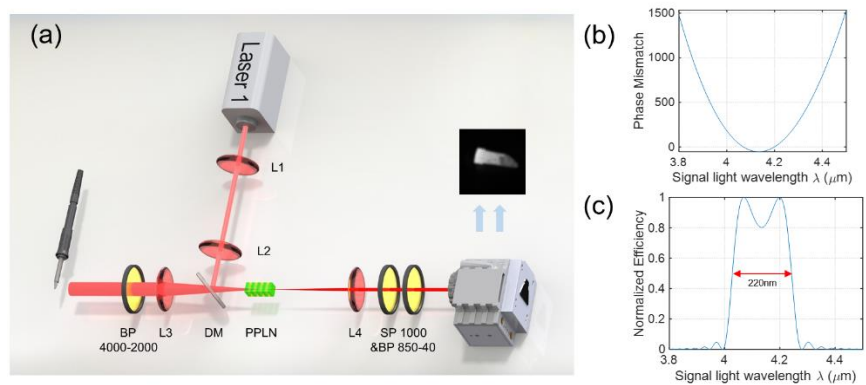

Fig. 1. (a) Schematic diagram of the experimental setup. L terms: the lenses; DM: dichromatic mirror; BP: band-pass filter; PPLN: periodically poled lithium niobate crystal. (b) Theoretical calculation of phase mismatch within the crystal (co-linear case) MgO-PPLN crystal length is 30 mm, polarization

period is 23.7 μm, and temperature is controlled at 39 °C. The pump light wavelength is 1080 nm. (c) Normalized upper conversion efficiency. The conversion bandwidth is 220 nm near 4.14 μm.

In our experimental setup, an industrial laser was used as the pump source with a central wavelength of 1080 nm. After passing through the lenses L1 ($f$ = 150 mm) and L2 ($f$ = 75 mm), the pump light was scaled down to pass through the crystal. The mid-infrared source is a temperature-controlled soldering iron or a standard blackbody oven that emits broadband blackbody radiation, which was collected by the lens L3 ($f$ = 100 mm). The BP 4000-2000 allows only mid-infrared light centred at 4000nm to pass within a bandwidth of 2000nm, the later BP 850-40 is similar. The type-0 (ZZZ) PPLN crystal has a dimension of 2 mm × 3 mm × 30 mm, mounted in a homemade temperature-controlled furnace. The thermal radiation of the target was detected by an sCMOS camera (Dhyana95 V2, Tucsen) at a central wavelength of 856 nm in the upconverted image after passing through the frequency conversion process. The lenses L3 and L4 ($f$ = 100 mm) formed a 4-f imaging system with the center of the crystal corresponding to the focal plane position. The short-pass 1000 nm and band-pass 850-40 nm filters at the output filtered out the pump light and its resulting second harmonics, while shielding the ambient light as much as possible. The imaging results of the soldering iron tip at 430°C are also shown in Fig. 1(a), where the curved contour of the edge corresponds to the limit of the imaging field of view.

For the upconversion detector introduced above, we further explored its NETD, which is an important parameter characterizing its imaging capability. The thermal radiation target is considered as an ideal blackbody, and the incident flux received by the detector can be described as [31]:

$$\phi_{e\lambda} = M_{e\lambda} A_0 \frac{\Omega}{\pi} \tau(\lambda) \tag{1}$$

$$M_{e\lambda}(\lambda, T) = \frac{8\pi hc}{\lambda^5} \cdot \frac{1}{e^{\frac{hc}{\lambda kT}} - 1} \tag{2}$$

where $M_{e\lambda}$ is the spectral irradiance, given by Planck's formula for blackbody radiation i. e. Eq. 2; $h$ is Planck's constant; $T$ is the thermodynamic temperature; $k$ is Boltzmann's constant; $\lambda$ is the incident light wavelength; $A_0$ is the area of the radiation source; $\Omega$ is the stereo angle of the detector input window to the target; and $\tau$ is the transmittance of the radiation propagating in free space as well as within the detector optical system. The variation of the incident luminous flux with temperature is given by:

$$\frac{\partial \phi_{e\lambda}}{\partial T} = \frac{\partial M_{e\lambda}}{\partial T} \frac{\Omega}{\pi} A_0 \tau(\lambda) \tag{3}$$

The responsiveness of the system concerning the incident flux is defined as $R_0 = N_s / \phi_\lambda$, where $N_s$ is the detector signal reading and $\phi_\lambda$ is the incident flux. The value of the responsivity varies in different systems, which is an important factor causing the difference in NETD. Here we consider a special case when the signal is exactly equal to the noise. According to the definition of the equivalent noise power, we have $R_0 = N_n / NEP(\lambda)$, where $N_n$ is the detector noise reading and NEP is acronym for Noise Equivalent Power. Applying this to Eq. 3, the amount of noise can be introduced for discussion:

$$\frac{\partial(\delta N_s)}{\partial T} = \frac{\partial M_{e\lambda}}{\partial T}\frac{\Omega}{\pi}A_0\tau(\lambda)R_0(\lambda) = \frac{\Omega}{\pi}A_0\tau(\lambda)\frac{N_n}{NEP(\lambda)}\frac{\partial M_{e\lambda}}{\partial T} \quad (4)$$

The NEP remains an important factor of inquiry in the discussion of system NETD. For a mid-infrared detector based on frequency upconversion, the NEP expression can be written as follows based on previous research work [22]:

$$NEP = \frac{\sigma_R}{pr\sqrt{\Delta f}} = \frac{\sigma_R}{\frac{\eta_{up}\eta_{det}}{h\nu_{MIR}}\sqrt{\Delta f}} = \frac{\sigma_R h\nu_{MIR}}{\eta_{up}\eta_{det}\sqrt{\Delta f}} \quad (5)$$

where $\eta_{up}$ is the upconversion process quantum efficiency; $\eta_{det}$ is the CMOS camera quantum efficiency; $\Delta f$ is the noise bandwidth; $\sigma_R$ is the total readout noise; and $\nu_{MIR}$ is the mid-infrared optical frequency. In the upconversion frequency converter, the main sources of noise are the Stokes noise of the shortwave pump, crystal thermal noise, image photon noise, and detector readout noise. Since the pump wavelength and other experimental parameters are kept constant, the noise components other than photonic noise can be considered as quantities independent of the wavelength of the mid-infrared light being upconverted. The photon noise of the upconverted light is proportional to the square root of the light intensity. According to the black body radiation equation, the variation of the radiation flux with wavelength is small within the conversion window of our UCD. Furthermore, the conversion efficiency of the detector is essentially the same for each frequency component in this wavelength band. Based on all of the above, the final estimated maximum change in photon noise is about 10%, which can be treated approximately as a constant. Therefore, we can treat NEP as a frequency-independent quantity in Eq. 4 to simplify the subsequent integration operation. Applying the small signal approximation, we obtain:

$$\frac{\partial N_s}{\partial T} \cong \frac{\Delta N_s}{\Delta T} = \frac{\omega A_0 N_n}{\pi NEP}\int_0^\infty \frac{\partial M_{e\lambda}}{\partial T}\tau(\lambda)d\lambda \quad (6)$$

Then:

$$\frac{\Delta N_s}{N_n} = \Delta T\frac{\omega A_0}{\pi NEP}\int_0^\infty \frac{\partial M_{e\lambda}}{\partial T}\tau(\lambda)d\lambda \quad (7)$$

By definition, NETD is equal to the change $\Delta T$ when $\Delta N_s/N_n = 1$ is satisfied, which gives:

$$NETD = \frac{\pi NEP}{\Omega A_0 \int_0^\infty \frac{\partial M_{e\lambda}}{\partial T}\tau(\lambda)d\lambda} = \frac{\pi NEP}{\Omega A_0\tau\int_0^\infty \frac{\partial M_{e\lambda}}{\partial T}d\lambda} = \frac{\pi NEP}{\Omega A_0\tau\frac{hc}{kT_B^2}\int_{\lambda_1}^{\lambda_2}\frac{M_{e\lambda}(T_B)}{\lambda}d\lambda} \quad (8)$$

$\lambda_1$ and $\lambda_2$ are the lower and upper limits of the detector conversion at mid-infrared wavelengths, respectively. In the present experiment, the quantum efficiency of the upconversion and detection processes can be written as

$$\eta_{up} = \frac{P_{up}/h\nu_{up}}{P_{MIR}/h\nu_{MIR}} = \frac{P_{up}\nu_{MIR}}{P_{MIR}\nu_{up}} = \frac{P_{up}\nu_{MIR}}{\int_{\lambda_1}^{\lambda_2}\phi_{e\lambda}d\lambda \nu_{up}} \quad (9)$$

$$\eta_{det} = \frac{P_{det}}{P_{up}} = \frac{\overline{x}_R h\nu_{up}}{P_{up}} \quad (10)$$

Where $\overline{x}_R$ is the average reading of the detector. Applying the above results to Eq. 8, we end up with:

$$NETD = \frac{\sigma_R k T_B^2 \int_{\lambda_1}^{\lambda_2} M_{e\lambda}(T_B) d\lambda}{\bar{x}_R hc \int_{\lambda_1}^{\lambda_2} \frac{M_{e\lambda}(T_B)}{\lambda} d\lambda} \tag{11}$$

Based on Eq. 8 and Eq. 11, the NETD of the upconverted detector can be directly estimated when its NEP is known, while for a newly built device, the NETD value can also be easily calculated based on experimental measurements. To test our calculations, we used a standard blackbody source instead of the soldering iron as the infrared target, which has better temperature stability. The imaging results are shown in Fig. 2(a), and the test area is the part of the figure inside the red box. The mean and standard deviation of the readings for each pixel were determined by taking several video frames at a constant temperature. We bring the results obtained in the experiment into Eq. 11 to obtain the first NETD matrix shown in Fig. 2(b). In the experiment, the temperature of the blackbody was set to 95°C, the pump laser power was 60W, and the single exposure time was 1s.

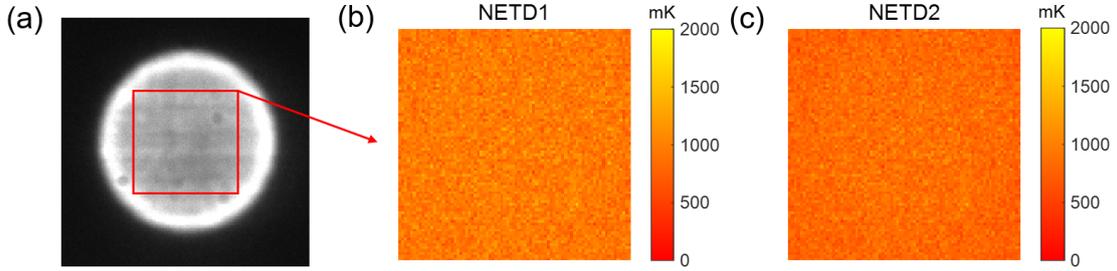

Fig. 2. (a) Results of the images taken in the NETD test. (b) NETD matrix given by calculation based on Eq. 11. (c) NETD matrix obtained based on the conventional measurement method.

To verify the accuracy of our evaluation method, the device was next tested using conventional NETD test means. The target temperature was first set to 90°C and 32 frames of the test image were obtained. The average of the grayscale values of each pixel in these frames was calculated to obtain the detector responsiveness matrix at this temperature. In the second step, the target temperature was set to 100°C and the above operation was repeated, obtaining a second set of responsivity matrices. The two are subtracted and divided by the temperature difference to calculate the gradient value of the responsivity concerning temperature. Finally, setting the target temperature to 95°C, 32 consecutive frames of data are acquired and the standard deviation of each pixel is calculated separately, which constitutes the noise matrix. Using the obtained results divided by the previous gradient matrix, the NETD matrix was obtained as shown in Fig. 2(c). The average values of the two NETD matrices in Fig. 2(b) and Fig. 2(c) are 951 mK and 829 mK. Although the feasibility of our evaluation scheme is initially demonstrated, the measured values of NETD are not satisfactory. This is because the upconversion detector does not operate in the optimal band due to the limitation of the operating temperature range of the blackbody source, a detailed discussion on this point will be given later. In addition, the standard deviation of each pixel point in the experiment is not entirely due to noise. Any perturbation in the experiment may cause ups and downs in the pixel grayscale values. Therefore, we chose a small region containing multiple pixels and reran the experiment with their total counts as samples. This is a means of compromise, which works well in practical measurements when the temperature target occupies more than one pixel. The final results obtained are 56mK and 48mK, respectively, which is an order of magnitude improvement compared to the case of a single pixel point.

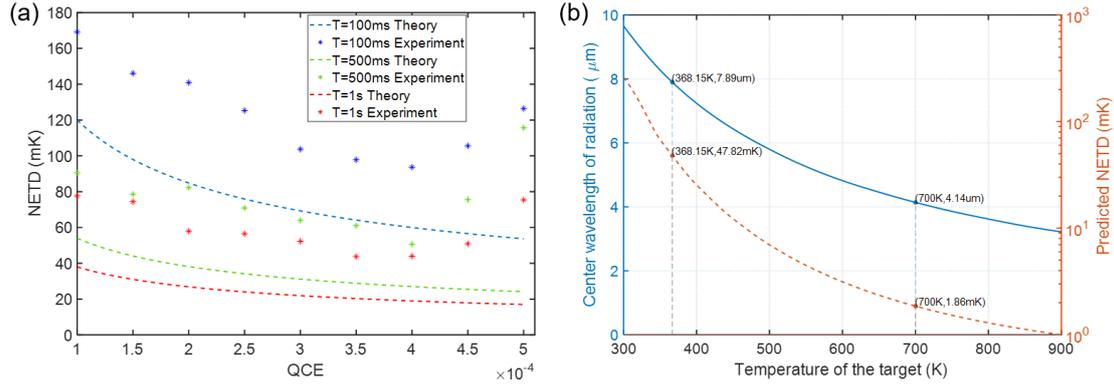

Fig. 3. (a) Comparison of measured and calculated NETDs as a function of quantum conversion efficiency (QCE) with different exposure times. (b) Central wavelengths of blackbody radiation at different temperatures (solid line) and corresponding NETD predictions (dashed line)

According to the theoretical analysis above, the NETD of the upconversion detector depends on various factors. In Fig. 3(a), the measured NETDs at different upconversion efficiencies and exposure times are shown separately and compared with the results calculated theoretically. The time T in the Fig. 3(a) legend is the single-frame exposure time, and completion of a set of tests requires completion of all 32 frames of image acquisition. The UCD quantum efficiency and pump light intensity were previously measured. The conversion efficiency was tuned by varying the pump light intensity during the experiment. The measured NETDs are overall higher than the theoretical predictions but are largely consistent with the trend shown by the latter. This deviation is reasonable considering the effect of the power stability of the pump laser (±1%) and the temperature stability of the blackbody source (±20 mK, given by the manufacturer). As the perturbation of both is random, it introduces additional noise into the experiment and makes the NETD higher than the theoretical value. It is worth noting that at higher quantum efficiencies the experimental results start to show some outliers. This is probably due to the fact that the corresponding pumped optical power exceeds 80 W at this point and the higher power density affects the thermal stability of the crystal. The theoretical prediction of the effect of the target centre temperature on the detector NETD is given in Fig. 3(b) and this result is calculated based on the same conversion bandwidth. The first set of results in the figure at 368.15 K (95 °C) is given experimentally, and this is combined with Eq. 11 to obtain the entire predicted curve. As can be seen from the blackbody radiation-temperature curve given by the solid line in the figure, the central conversion wavelength of the detector in this experiment is 4.14 μm and the best matching temperature is ~700 K. As mentioned above, the upconversion detector does not operate in the optimal temperature interval. For our experimental setup, the NETD using a target center temperature of 700 K is only about one-twentieth of that at 368 K. Borrowing the results given in Fig. 2(b) and (c), the predicted mean value of NETD for each pixel is about 37 mK under the same experimental conditions using a 700 K target. Meanwhile, for the case where the target area occupies multiple pixels, the NETD prediction drops from 48mK to about 2mK.

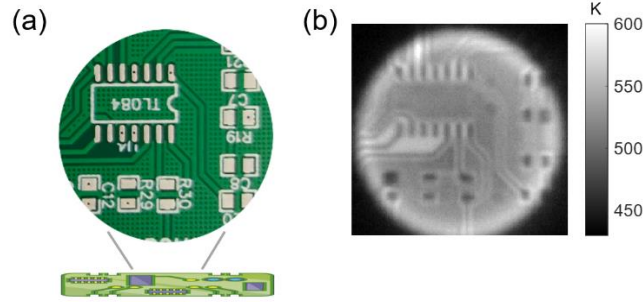

Fig. 4. (a) The PCB board used for the experimental tests. (b) Photographic results of the upconversion detector (part of the area). The PCB board was preheated

In Fig. 4, we show the results of the upconversion detector for a general temperature object. An ordinary PCB circuit board was laminated to the electric heating platform for ten seconds. Due to the different thermal conductivity of the substrate, the internal copper conductive layer, and the metal part of the surface, a clear temperature distribution is presented. With the help of the pre-calibrated results, we can give the specific temperature distribution of the target, as shown in Fig. 4(b). The internal structure of the PCB board is simply presented through heat conduction. This technique could have potential applications in engineering fields such as non-destructive testing of objects. For example, impurities, internal damage, and cavities can cause changes in thermal conductivity so that the target presents a distinguishable temperature distribution after uniform heating.

The best NETD test result obtained in the experiment was ~50 mK, while in the ideal case the theoretical prediction of the NETD for our device is about 2 mK, which is significantly lower than the NETD (25 mK) of the current thermal imager used for the study (FLIR SC7000). The inability of the blackbody furnace used in the tests to operate at higher temperatures has become a major factor limiting the NETD optimization of the system at this time. Without considering the objective experimental constraints, the theoretical analysis above provides a viable option to further reduce the NETD. Eq. 5 and Eq. 8 show that the NEP and NETD of the system can be optimized by increasing the upconversion quantum efficiency and the measurement time. And the conversion bandwidth of the system and NETD are approximately inversely related while keeping the NEP constant. Within the operating window of our UCD, the phase matching condition is always satisfied and therefore a broadband conversion will result in a higher overall conversion efficiency compared to a narrowband UCD. In this case, since the thermal noise of the pump, the readout noise of the detector is constant, a larger converted signal strength in broadband conditions will lead to an increase in the signal-to-noise ratio, i.e. a decrease in NETD.

In summary, we investigated a MIR detector based on a frequency upconversion process and its NETD characterization. The conversion of MIR radiation with ~220 nm bandwidth was achieved using a single QPM crystal. We theoretically analyzed the factors influencing the NETD of the system, gave a calculation equation to estimate this metric, and experimentally compared the results with conventional NETD tests. We also study the test results of the device for general temperature targets. The optimal detection temperature of our detector is around 700 K. The choice of different upconversion crystals allows the migration of the detection window. The present work provides a reliable NETD evaluation scheme for such frequency-based upconversion IR imaging systems. At the same time, our device demonstrates good temperature detection performance, which has potential applications in remote sensing, material structure detection, and etc.

**Acknowledgments.** We would like to acknowledge the support from the National Natural Science Foundation of China (NSFC) (11934013, 92065101); Anhui Initiative In Quantum Information Technologies (AHY020200); Innovation Program for Quantum Science and Technology (2021ZD0301100); National Key Research and Development Program of China (2022YFB3607700).